# Electronic administration in Spain:
# From its beginnings to the present

Antonio Muñoz-Cañavate [a], Pedro Hípola [b]

**Abstract:**

This study presents the basic lines of electronic administration in Spain. The complexity of the Spanish political-administrative system makes such a study challenging, in view of the considerable degree of autonomy and competences of the regional administrative bodies and local agencies with respect to the central government, the former being more visible in the 17 regions of Spain. Nonetheless, the central government maintains a series of legal instruments that allow a certain common framework of action to be imposed, aside from what is put into effect through diverse programs aimed precisely to develop common tools for the regions and municipalities of Spain. After an introduction that provides some necessary background, this study describes the legislative framework in which Spain's electronic administrative system has developed. The data included in the study refer to investment in information and communication technologies (ICT) and the services offered by the different Administrations on the internet; internet access by citizens, homes, businesses, and employees, as well as the interactivity existing with administrations by means of the internet; the origins and rise of various political initiatives of the Central Government involving electronic administration; and finally, the situation of civil service personnel, as catalysts of the success of Information Society in the Public Administration within Spain.

[a] Departamento de Información y Comunicación, Facultad de Biblioteconomía y Documentación, Universidad de Extremadura, 06071 Badajoz, Spain

[b] Departamento de Biblioteconomía y Documentación, Facultad de Comunicación y Documentación, Universidad de Granada, 28071 Granada, Spain

# 1. Introduction

This article provides an overview of e-administration in Spain, from the beginning to the present, based on studies carried out within Spain's public administration, the university, and by firms and foundations in sectors such as telecommunications (e. g. Telefónica and France Telecom). The second section looks at the pertinent legislative and normative framework, and the third section focuses on ICT investment and services in operation in Spain. The fourth part of the study describes the level of access by citizens and businesses to internet and services that allow for interaction with public administrations, based on data gathered by Eurostat. The fifth section reviews the most relevant political initiatives of the GSA since 1992; and finally, the sixth part describes the role of public employees and their perception of e-administration, as expressed through surveys by Spain's Center of Sociological Research.

## 1.1. The political-administrative structure of Spain

When the political-administrative situation of Spain is considered, the complex context of its origins in the present democratic government, established in 1977, and in the ensuing Spanish Constitution of 1978 must first be described. The year 1975 was marked by the death of General Francisco Franco, who had ruled Spain since 1939 under a dictatorship; at this point a period of transition toward democracy was initiated –a process internationally praised as exemplary– which was made possible by a pact on the part of Spain's different political parties and their ability to overcome their differences in order to aspire to a common goal.

Yet present-day Spain also stands as a country that has experienced tremendous economic growth over the past three decades, and has witnessed the arrival of a wave of immigration from Latin America, Africa, and Eastern Europe (most notably), which has made the population increase considerably. Records document that there were 46 million inhabitants in Spain in 2009.[1]

The Spanish Constitution of 1978 transformed a deeply centralized State into one that was decentralized by organizing it into regions or autonomous communities. Clarification of this situation, albeit briefly, is essential for an understanding of how the administrative panorama of Spain has developed –and therefore of the plans involving electronic administration– as these Administrations have great leverage in the management of their separate competences and interests, and each may be governed by norms of their own, though with some limited degree of coordination among them all.

Thus, Spain at date has a General Administration of the State or *Administración General del Estado* (AGE), with central and peripheral structures, led by the President of the Government and his Ministries. On a second tier we have the regional Administration, with 17 autonomous communities enjoying a very high regime of exclusive competences, and two autonomous cities (in northern Africa), Ceuta, and Melilla. The regions are governed by respective regional governments. In turn, these regions may be comprised of a single province or a number of provinces.

Next, there is the local Administration. When there are several provinces involved, we find yet another administrative level: the *Diputación* of the province (there are 41 of these). However, in the island regions of Spain (Canary Islands and Balearic Islands) these are not called *Diputaciones*. In the Canary Islands they are known as *Cabildos*, and in the Balearic Islands they are called *Consells* (ten in all). Meanwhile, the *Diputaciones*, *Cabildos*, and *Consells* have their own political bodies. Finally, the provinces contain municipalities governed by City or Town Halls—the administrative unit closest to the citizen. In Spain there are 8112 Town Halls.

The panorama described here reflects the great complexity of the Spanish state (see Table 1) as well as that of the different policies in effect nationwide. Such is the case for administrative modernization projects. The central government, the regional governments of the autonomous communities, the governments of the provincial Diputaciones the governments of the insular Cabildos and Consells, and those responsible for Town Halls across Spain have, since the mid-nineties, established policies aimed to bring their administrative services closer to the citizens.

The central government wields a set of specific competencies in the following categories: defense and public security; international relations; the administration of justice; mercantile, penal, labor, and civil legislations; and the planning of economic activity or the infrastructures that are found in more than one autonomous community. However, the practical transfer of the remaining competencies to the regional governments leaves the citizen with the impression that what the regional or local government does –or does not do– will affect the citizen's life more that what the national government might.

**Table 1**
The Public Administration and its territorial realm in 2008.

| Territory | Administration | Units |
|---|---|---|
| Spain | General state administration | 1 Central Government with 17 ministries |
| Autonomous Community[a] | Regional administration | 17 regional governments and 2 autonomous cities |
| Provinces and Islands | Local Administration: *diputaciones* | 41 |
| | Local Administration: *consells* (Balearic Islands) | 3 |
| | Local Administration: *cabildos* (Canary Islands) | 7 |
| Municipality | Local Administration (city/town halls) | 1,112 |

[a] The cities of Ceuta and Melilla, in northern Africa, have an autonomous structure of their own, officially becoming *ciudades autónomas* in 1995, with a regime similar to that of the *Comunidades Autónomas*.

This great political-administrative maze often gives rise to problems rooted in the decision-making processes or political workings of some administrations as opposed to others—problems that can be aggravated when the administrations are governed by rivaling political parties.

Although the Spanish Constitution of 1978 and posterior laws acknowledge the competencies of the State by means of the "bases of the legal regime of the Public Administration," and of the "common administrative procedure," which allow for necessary coordination among different governments, it is no less certain that the political struggles of a country like Spain, with its strong nationalistic movements, have led to disputes generating inefficiency, and even stupor among the body of citizens. The "lack of coordination" is noted in many of the reports published in Spain and makes known the reality of the public administration, to such a degree that the initial model of the development of electronic administration has come to be defined as one of "islands" of implantation, which should advance toward collaboration in order to be functional.

The relationship between and among administrations requires the creation of some coordinative structures. It is in the State Administration where the coordination is most evident in the sense that it is governed by a single political power, unlike that of the regional administration or the Town Halls. Hence, in the central administration, the Ministry of Public Administrations is the department in charge of modernization in the implementation of information and communication technologies (ICT), but we also have a Higher Council for Electronic Administration (*Consejo Superior de Administración Electrónica*) and a General Direction of Administrative Modernization (*Dirección General de Modernización Administrativa*).

Each autonomous community or region has its own mechanisms for internal coordination, but among the 17 separate communities and the two autonomous cities there exists in the central government a sectorial conference and a technical committee for matters of electronic Administration.

Such complexity may dissuade one from embarking on in-depth studies concerning electronic administration. Indeed, political reticence and friction among regional governments can reduce the efficacy and reliability of those instruments of coordination that the regions have set up (no regional government wants to appear to be below the other). As a result, the best assessment tools, whether in the framework of e-government action or of e-democracy, must be sought in independent entities, such as the annual reports put out by the Orange Foundation of France Telecom. For example, the first coordinating initiative among all the regions was set in motion in 2005 thanks to the *Comité Sectorial de Administración Electrónica*, which involved regional and central level government entities. As implied previously, such results cannot go unchallenged; the regional governments themselves were answering a questionnaire that was agreed upon by all. Moreover, in the different editions of this survey done to date, some governments did not present their results.[2]

The local administration is even more complex. While in each province the provincial disputations normally dispose of mechanisms for the coordination of issues, there are also municipal associations that collaborate on common policies. The most important association of this sort in Spain is the Spanish Federation of Municipalities and Provinces (*Federación Española de Municipios y Provincias*, or FEMP) which encompasses 85% of the local entities.

*1.2. The processes of modernization and electronic Administration*

We have thus far described the complexity of Spain insofar as the different layers of government and administration, but what about the processes of modernization? Over the past 15 years, citizens of Spain, as in the rest of the world, have grown familiar with a new form of interaction with government agencies that is known as electronic administration. The incarnation of e-administration has obliged governments to make laws in this direction and establish means of interaction with citizens using ICT. Studies worldwide are many and have looked at the impact of electronic Administration in other countries (Allen, Juillet, Paquet, & Roy, 2001; Chan, Lau, & Pan, 2008; Chen, Pan, Zhang, Wei Huang, & Zhu, 2009; Gauld, Gray, & McComb, 2009; Groznik & Trkman, 2009; Holliday & Yep, 2005) such as those that underline the benefits of e-government and its bases (Gupta & Jana, 2003; Jaeger & Thompson, 2003), or those analyzing the different stages of evolution through which the processes of this new paradigm have passed (Gil-García & Martínez-Moyano, 2007; Hiller & Bélanger, 2001; Layne & Lee, 2000), the costs involved (Bertot Jaeger, 2008) or its development in different countries (Esteves & Joseph, 2008; Mitra & Gupta, 2008). In the specific case of Spain, there are studies exploring administration-related information policies (Cornella, 1998), legislation and its relationship with document processing and electronic administration (Bustelo & García-Morales, 2008), and the work-related implications of e-government for information science professionals (Chain Navarro, Muñoz Cañavate, Salido Martínez, 2008b). These are just a few examples.

It is not our intention to underestimate technological innovations such as the videotex, although aside from its implementation in France (Moulaison, 2004), it has known little success; or the client-server application of the internet "gopher," whose short life span hardly allowed for its development. We could, however, briefly describe the scenario preceding the surge of the World Wide Web in the early 1990s.

In the western world, a change in the means of acting and working in public bodies had been foreseen since the middle of the past century, only to be accentuated by the world economic crisis, a crisis that was provoked by soaring oil prices in the 1970s. The consequent need to reduce public budgets nourished a search for new formulas that would make for greater efficiency in the public workplace. This scenario saw the arrival of microcomputing in the 1980s, making it possible to turn endless manual tasks into automatic processes, which encouraged political bodies to implement and foment modernization processes.

The development of the internet, and of other networks which would eventually fuse with microcomputers, gave us an ideal platform for citizen interaction. The fluxes of diversified and increasing information among citizens, businesses, and the administration led to enhanced relations within the bureaucratic structures themselves. The foundations for change in work standards had been set within the Public Administrations, both in back-office and front-office processes.

Political pressure can also be seen as an important motor of development. From the eighties onward, more and more initiatives have been directed at generating the projects of electronic administration. In February of 1995, Brussels was the seat of an international conference on the information society, with the G7 and the European Commission defining eleven projects. One of these projects was "Government online," which was coordinated in the early stages by the United Kingdom's Central Telecommunications Agency and Canada's Secretary of the Treasury. This international conference, with political precedents in different world regions, marked a turning point. From then on, political strategies dedicated to propelling an information society in the context of strong technological advances were accelerated. In Europe, for example, a series of pluri-annual plans appears: eEurope 2002; eEurope 2005; and eEurope 2010.

Furthermore, other international organisms such as the OECD have spoken out clearly on the benefits of the incorporation of electronic administration to the lives of citizens in general. The publishing of a document in 2003 and titled "The e-Government Imperative" confirms this. Likewise, in 2004 a document called "Electronic Governance"[3] reflected the European Council's insistence on the need to delve into technical matters as well as democratic governance.

E-administration thus became a new paradigm by which numerous conceptual approaches were accommodated. Our study focuses on one contribution formulated by the Commission of European Communities: "the new technologies and communications can help the public administrations face a number of challenges. However, the emphasis should not be placed on the ICTs themselves, but on their combined use with organizational changes and with new aptitudes destined to improve public services, democratic processes and public policy. This is what electronic administration (e-government) refers to."[4]

A law passed in Spain in 2007 regarding *Acceso Electrónico de los Ciudadanos a los Servicios Públicos* (Electronic access by citizens to public services) is a good example of what electronic administration means in terms of the parameters of space and time. In expounding its motivating forces, the law states that

> the greater proximity of the citizen to administration deriving from the autonomous and local decentralization has not managed to overcome the barriers that continue to distance the citizen from the administration, from any administration, including the central one, and which is often nothing more than the barrier that time and space erect: the time required for dedication to this relationship for the realization of many bureaucratic tasks of daily life that sometimes stem from the need for initial information that calls for an initial physical presence, plus the further trips and time dedicated to later tasks involving the administration for the most elementary activities […] at any rate, those first barriers in relations with the administration –the distance traveled and the time required– are not necessary nowadays. The information and communication technologies make it possible to bring the administration into the living rooms of citizens or to the offices of professionals and businesses (BOE, 2007a).

In short, a process of change in the interactions of citizens and firms with their public services was generated, and the transformation of the public administration from heavy bureaucracy to more flexible structures was a new goal. At the same time, administrative interactions among the different levels of governments were transformed.

This also stood as a new basis for enhanced decision-making and greater informative transparency within a strictly political frame-work. In the mid-twentieth century, John Gauss foresaw this new relationship between public administration and political science. We have entered a new conceptual realm known as e-democracy. According to Arsntein (1971), new forms of public participation and interaction with the political sphere would bring the other realm of Administration into conformity, on five concrete levels that involve the possibility of: being informed (information); creating a bidirectional relationship with governments (communication); developing consultation to improve awareness of public opinion (referenda, surveys, polls); establishing processes of evaluation and reflection about any topic (deliberation); and participating in decisions and elections (Arsntein, 1971).

Electronic administration can also provide for the introduction of tools that make it easier to fight corruption. Spain has been associated with numerous cases of building-related corruption in recent years, under the umbrella of huge economic growth in the construction industry. Some of these cases have landed politicians in prison. One example of way in which eadministration can fight such corruption is the online publication of plans for the development of a number of Spanish municipalities. Since 2007, this pilot project has allowed the intended building projects involving ten municipalities to be made public via the internet.[5]

What's more, the use of e-administrative services means great monetary savings. The actual reduction in public spending is just now beginning to be quantified.[6] The plan for action of the electronic administration i2010 estimates the annual budgetary savings in the European Union on the whole to be at 50 billion euros as a result of the implementation of e-administration. The same document quotes savings from the electronic billing to taxpayers in Denmark at 150 million euros per year, and 50 million for businesses (Comisión Europea, 2006).

Within Spain, the discontinuation of the paper version of the Official State Bulletin (*Boletín Oficial del Estado*) means a savings of 6.3 million euros and the rescue of 3,500 tons of paper previously consumed every year, whereas the cost of renovating the systems, certifying the electronic signature, the servers and the new system of production comes to less than 200 thousand euros (Reventos, 2008).

## 2. Spanish regulation and e-administration

Obviously, Spain's laws have had to be adapted to the new setting of electronic administration, thereby allowing its relations with citizens and businesses to be regulated. This section offers an overview of the corpus of norms and laws underlying the legal foundations for the development of e-administration in Spain.

As pointed out earlier, the arrival of Democracy in Spain's first general elections of 1977 began a regimen of personal liberties that would also be made explicit in the text of the Constitution of 1978. Among its principles, the Spanish Constitution defines the right to information[7] (article 20.1), the effectiveness of the Administration (article 103), and citizens' access to government archives and administrative records (article 105). Taking advantage of the move toward decentralized politics, from this point forward, the administration of Spain initiated a full-fledged process of legislative and political modernization.

The great number of norms appearing in Spain over the past few years, whose objective is the adaptation of the new processes of interaction among administrations, citizens, and businesses in the legal framework of the new administration, arose from the legislative and executive powers of the Spanish State, and also from the need to incorporate within the judicial order the different directives of the European Union.

We might say that there is unanimous agreement in Spain as to chronologically synchronizing the onset of legal transformations targeted at an electronic administration, and the countdown comes in 1992. This year represents the beginning of the profound transformation of Spain's administrative bodies. Though such initiatives had been developing since the 1980s in successive attempts to modernize processes involving communication with citizens, and the modes and methods of internal functions,[8] the so-called Law of Juridical Regime of Public Administrations and Common Administrative Procedure (*Ley de Régimen Jurídico de las Administraciones Públicas y del Procedimiento Administrativo Común*) of 1992 (BOE, 1992) (to be applied in all the Public Administrations of the State) marks the beginning of the end of legal problems arising in Spain as a result of applying information technologies to administrative procedures.[9]

Cited most often are three articles of this Law: Article 45, dedicated to the incorporation of technical media; Article 38, referring to the computerization of registers and archives; and Article 59, regarding the notification that could be made by any means, although in Article 59 a law regarding fiscal measures from 2001 (*Ley 24/2001, 27 December, Medidas Fiscales, Administrativas y del Orden Social*) will allow for telematic means to be used if the interested party has so indicated (BOE, 2001). Finally, the General Tax Law (*Ley General Tributaria*) of 2003 authorizes automatic administrative action aside from the electronic reproduction of documents.

In Article 45 of the aforementioned Law of 1992, explicit reference is made to the use of electronic, computer, and telematic means in the exercise of administrative action. It also makes reference to the validity of electronic documents, which would "enjoy the validity and efficacy of the original document, as long as their authenticity, integrity and conservation are guaranteed."

Various provisions would later stem from the Law of 1992. Such is the case of the Royal Decree (*Real Decreto*) 263/1996 (BOE, 1996) to regulate the use of electronic, computer and telematic techniques by the General Administration of the State, expounded in article 45 of the 1992 law, with a well-defined objective: to eliminate fears surrounding the use of new technologies in any of the activities of the Administration.

In 2003, another subsequent decree (R.D. 209/2003) led to the regulation of records, telematic notifications, and certificates and transmissions. Accordingly, it is stated that any interested party "making manifest his/her will to be notified by telematic media in any procedure whatsoever should have available, under the conditions established, an address suited to such a purpose, which should be unique for all possible notifications to be practiced by the General Administration of the State and its public organisms" (BOE, 2003a). This service is available in Spain thanks to the collaboration between the Ministry of Public Administrations and the postal service company (*Correos*) which makes available to any requesting such the possibility of receiving by telematic means the notifications that are currently received on paper.[10]

One of the very most effective innovations for the implementation of electronic administration, however, is the electronic signature, which ensures the validity and the confidentiality of citizens' electronic transactions with the

Administration. The European directive 1999/93/CE sets forth a community framework for the electronic signature that gave rise in Spain to the Royal Decree of electronic signature (*Real Decreto Ley 14/1999 de firma electrónica*).

Security in this type of communication is provided through services of certification that ensure the identity of the user.[11] Its later development as a Law, in view of accumulated experience, causes a new and more complete normative to be enacted in 2003: Law 59/2003 (BOE, 2003c), to highlight the Electronic Identity Card as an electronic certificate of full validity. This document accredits the identity of its holder and permits the electronic signing of documents.[12]

Among the novelties of the text of 2003 with respect to the text of 1999, we find the very denomination of electronic signature deemed equivalent to the handwritten signature. Hence: "the acknowledged electronic signature will have, with respect to the data given in electronic form, the same value as the handwritten signature in relation with those given on paper."

The Law 34/2002, or Law of Services of Information Society and Electronic Commerce (*Ley de Servicios de la Sociedad de la Información y del Comercio Electrónico*) (BOE, 2002), comes into force on July 11, 2002. In essence, this norm is intended to heighten entrepreneurial efficiency, as it includes several European directives related to the domestic market, the protection of consumer interests, and the broad development of the networks of telecommunications with the increased possibility of choice on the part of users. It also affects the Administrations in diverse ways, such as contracting services by electronic means, and accessibility for persons with a handicap or advanced age. As one of the additional provisions, it mentions the necessary adaptation of the pages of Public Administrations to comply with acknowledged criteria for accessibility.

Among the legal norms borne in Spain to solve the problem of accessibility for handicapped persons, we also have the "law of equal opportunity, non-discrimination and universal accessibility of handicapped persons" (*Ley 51/2003, 2 December, de igualdad de oportunidades, no discriminación y accesibilidad universal de las personas con discapacidad*) (BOE, 2003b). Its seventh and final provision obligates the central Government to fix basic conditions for the non-discriminatory accessibility for the use of technologies, products and services related with the information society and any means of social communication. This translates in 2007 into the Royal Decree 1494/ 2007, on 12 November, approving the Regulation on basic conditions for the access of persons with a handicap to technologies, products and services related with information society and means of social communication (Real Decreto 1494/2007, de 12 de noviembre, por el que se aprueba el Reglamento sobre las condiciones básicas para el acceso de las personas con discapacidad a las tecnologías, productos y servicios relacionados con la sociedad de la información y medios de comunicación social) (BOE, 2007b), and which makes explicit mention of the principals of the Web Accessibility Initiative (WAI) of the World Wide Web Consortium.[13]

Yet if there were a single law that had done more than any other for the effective development of the e-administration in Spain, that would be the Law of Electronic Access of Citizens to Public Services (*Ley de Acceso Electrónico de los Ciudadanos a los Servicios Públicos*), in Spanish LAECSP, approved in June of 2007 (BOE, 2007a). The novelty of this law resides in its consecration of the right of citizens to communicate with the administrations by electronic means, which obliges the latter to have the electronic and telematic means so that citizens can gain access to information and services, present requests, applications or appeals, make payment, or receive notifications or communications from the Administration. The former option of the administrations to set up such mechanisms or not thereby became the obligation to do so. In sum, "may be lent" became "should be lent."

The following sections structure the major changes brought about by the new law.

(a) Citizens' rights are regulated through electronic media. Here, the obligation of the administrations to make available different channels or means for lending electronic services methods such as the mobile telephone, cable television or any other technological innovation that could appear in the future.

(b) The Public Administrations are obliged to facilitate to other administrations the data on those citizens who consent to such. In this way, the administrations do not have to request documents from citizens that are already in their power, for example, photocopies of the national identity document (DNI).

(c) The figure of the User Defender is created, oriented to attend to complaints and act upon suggestions and proposals for improving

the relationships of citizens with the Public Administrations through electronic media.
(d) An electronic headquarters is regulated as an electronic address that is to be managed by an Administration, which should oversee its integrity, veracity and the updating of the information and the services it has.
(e) Also established are the forms of identification and authentication of the citizens and of the administrative organisms themselves. Thus, the electronic National Identity Document is established as a formula for extending the general use of the electronic signature, and all administrations are obligated to admit electronic certificates recognized in the realm of the Law of electronic signature.
(f) The digital gap and digital illiteracy are contemplated in the Law. The civil servants can accredit the will of citizens to have electronic access to administration.
(g) The registers, communications, and telematic notifications are regulated, but increase the possibility of the presentation of applications or requests to administrations. Also established are criteria for the validity of electronic documents and archives, or of electronic copies.
(h) The complete electronic management of the administrative procedures using electronic media is defined. Thus, the entire process of initiation, instruction, and termination of procedures by electronic means is regulated, including the obligation of the administration to provide information about the processing status of procedures that affect an individual.
(i) In the end, the mechanisms of cooperation between the State administration with the regional administrations and with the local administration are set forth. This organ, known as the Sectorial Committee of e-administration, ensures compatibility across systems used by the different administrations, aside from preparing joint programs. To a certain extent, the aim is to fix legal mechanisms above and beyond territorial tensions between regional governments of nationalistic markings and the State government, which arise now and then.

Nonetheless, Spain has not yet solved one particular problem that many civil groups express when they call for a law for access to public information, which does not yet exist and that would, at least in part, limit serious deficiencies and corruption, perceived by citizens to be severe weaknesses of the political system. According to a barometer elaborated by Transparency International since 2003, for most Spaniards, the most corrupt sector of society is that of the political parties.[15]

This demand for a law controlling access to public information was made manifest at the end of 2007, at the celebration of the *International Right to Know Day*, on September 28, when fifteen Spanish entities and NGOs denounced the government of Spain for not fulfilling its electoral promise of adopting a law of access to this type of information. In the opinion of this collective of voices, "all citizens have the right to know how decisions are made within any public administration, and to know how taxpayers' money is spent. The best tool for guaranteeing full recognition and exercise of this right is the existence of a law specifically regarding access to information."[16]

Traditionally, Spain has not been a country in which the public administrations were characterized by transparency. Many more reasons can be cited to explain this situation. Alfons Cornella points out that while Spaniards are quite communicative, organizations here tend to be opaque, both in the public and in the private sector. Cornella identifies the lack of an "information culture" in Spanish society as one probable reason; partly due to the existence of an educational system that is more concerned with imparting knowledge that will be of lifelong use than in teaching informational skills that can be used to continually update one's knowledge (Cornella, 1998).

To some extent, Spain's public administration moves in semi-darkness in the political realm, which we might observe, for instance, in the continually poor results related to decision-making. The table presented below (Table 2) shows the results of a study (Chain Navarro, Muñoz Cañavate, & Más Bleda, 2008a) applied in early 2008 to the websites of the 52 city halls that represent the capitals of provinces across Spain, plus the two autonomous cities of Ceuta and Melilla. The purpose of the study was to assess the contents and services offered on the websites of a sample of Town Halls around Spain; in general, the study reflected that there was a scarce disposition on the local level to offer information that was of greatest interest to citizens. Although the municipal ordinances appear in 86% of the websites analyzed, other items identified as critical appeared far less frequently. For example, in many areas, a citizen cannot get in contact with his political representatives (mayor or councillors) through e-mail, and only a very few city or town halls (5.77%) broadcast municipal plenary sessions over the internet. Another example is that there are no formulas for permitting citizens to vote by internet, despite the fact that there are technological applications that would allow for such citizen participation.

**Table 2**
Documents of political participation (%).

| | |
|---|---|
| Acts of plenary sessions | 48.08 |
| Municipal budgets | 51.92 |
| Municipal information bulletin | 42.31 |
| Decrees and/or edicts from Mayor's office | 86.54 |
| E-mail contact with the mayor | 46.15 |
| E-mail contact with city council | 30.77 |
| Retransmission of municipal plenary sessions | 5.77 |

In this context we must, however, underline that while the LAECSP of June 2007 required Administrations to make telematic contact possible, electoral privileges were not reflected—the voting procedure is guided by an Organic Law of the General Electoral Regime (*Ley Orgánica de Régimen electoral general*). In Spain, the Undersecretary of the Interior is working on the Individual Platform of Electronic Voting (*Plataforma Individual del Voto Electrónico*, or PIVE), which tries to ensure that any citizen wishing to exercise their right to vote is guaranteed that technical security will prevail throughout the process (Martínez Domínguez & García de la Paz, 2007).

As described earlier, at least some impediments in accessing information would be related to the political-administrative complexity of Spain, since disputes regarding the transfer of diverse political functions and operations are common, and a lack of coordination may be the norm rather than the exception. Thus, informational lapses can be seen as one of the elements behind the slow growth of Spain's productivity over the past decade, and an obstacle to be overcome for future development. As underlined in the OCDE report of 2007 (OCDE, 2007), a greater transparency and coordination of the programs articulated by central and autonomous governments would do away with duplicity, which would in turn facilitate access to information for "Small and Medium-sized Enterprises" (in Spanish, PYMEs), and reduce the risk that these measures could become mere industrial policies favoring local businesses.

Another legislative contribution is Law 37/2007, dated November 16 (BOE, 2007c), regarding the re-utilization of information from the public sector. This law was designed by several ministries: Justice, Industry, Commerce and Tourism, Presidency and Public Administrations, at the urging of the Ministry of Culture. "Re-utilization" is understood here as the use of those documents that are in the hands of the Public Administrations and organisms of the public sector, and are used for both commercial and non-commercial purposes.

The preamble of this Law defines document in generic form as

> embracing all the modes of representation of acts, deeds or information, and any recompiling of the same, regardless of format (written on paper, stored in electronic form or as an audio, visual or audiovisual recording) conserved by the Administrations and organisms of the public sector; and it includes a negative delimitation of the scope of application, enumerating those documents or categories of documents that are not affected thereby, attending to diverse criteria (BOE, 2007c)

Excluded are documents where intellectual or industrial copyright applies.

Some of the cases denounced in the realm of intellectual property rights related to the Spanish Project known as *Hemeroteca Digital de la Biblioteca Nacional Española* (Digital Hemerotecque of the Spanish National Library). This project, which was targeted toward the consultation and public diffusion over internet of the Patrimony of the Library of Spain, as conserved in the *Biblioteca Nacional*, had been criticized for including the copyright sign of the library on material pertaining to the public domain. As a consequence, in the Hispanic Digital Library (BDH), inaugurated in January of 2008,[19] free and gratuitous access is given to some of the principal works of Spain's cultural heritage, with no mention of copyright or appearance of the copyright logo.

We end this section by mentioning the last of the great regulating laws of 2007, known as the Measures to Promote Information Society (*Ley 56/2007, 28 December, Medidas de Impulso de la Sociedad de la Información*) (BOE, 2007d), which partially modifies some of the previous norms, such as that of the Services of Information Society and Electronic Commerce (*Ley 34/2002 de Servicios de la Sociedad de la Información y del Comercio Electrónico*) or that of electronic signature (***Ley 59/2003 de firma electrónica***). It establishes the requirement of electronic invoicing or billing in cases of hiring services in the state public sector. In this way, the central government itself is obligated to encourage this type of billing in its dealings with small businesses and even "microbusinesses," to a common end: fomenting electronic commerce.[20]

## 3. Investment in ICT and e-administration services

This section takes a look at the most significant data regarding the three levels of administration (State, regional, and local) in terms of ICT, websites, and e-administration services currently functioning in Spain. There is, unfortunately, no single periodical report with a homogeneous methodology for all the administrations involved, for which reason we must cite different sources and papers.

In order to more clearly demonstrate the extent of adoption of ICTs in Spanish Administrative bodies, and therefore the reality of e-administration in Spain, we will look at: (a) expense in computer and telecommunications in the different Administrations, (b) the percentage of computers with access to internet and intranet, and (c) the actual services that Administrations lend through their websites. To support our analysis, we consider two primary types of sources: those that come from within the Administration, and those that analyze this reality from outside. The sources used to determine the situation and use of ICT and systems within the administration were, essentially: (a) the REINA

Report, containing data on the State Administration, (b) the IRIA Report of Informational Resources of the Public Administrations for the local governments and GSA, and finally, (c) the e-administration questionnaires (CAE), dealing with the administration of regional governments and published by the "Observatory of Electronic Administration", which has produced five reports from 2005 to 2009, though not all of the regions have participated consistently.

### 3.1. Expenses in ICT

In the State administration, expenses in computer-related services stood as 80% of the total ICT expenses, as opposed to 20% in telecommunications. In the local administrations, the corresponding figures are 79% for computer expenses and 21% in telecommunications (MAP, 2008).

A study by IDC España that reported on the inhibitors of ICT investment in public administrations pointed to different reasons for the lack of investment in ICT. Whereas for local administration the main reason was the budget, for the AGE the reasons included a lack of qualified personnel, a lack of political support, and the resistance to change. The regional administration underlined as inhibiting factors the internal resistance to change and a lack of budgetary resources (Fundación Telefonica, 2006). In a report brought out by the United Nations Network in Public Administration and Finance (UPAN), increased investment in ICT is acknowledged, albeit along with the following inhibitors that act as barriers to the effective development of e-administration in developed countries: institutional laziness rooted in insufficient planning for the introduction of change; the lack of qualified staff; insufficient funds for financing innovative projects; and problems deriving from the rapid evolution of the hardware and software, which makes them soon obsolete—problems that would be common to many countries (Fundación Telefonica, 2004).

### 3.2. Websites and services

Studies regarding the appearance of websites have been a constant development in recent years in Spain and address websites at all levels of Spanish government: State Administration (Muñoz Cañavate, 2003), regional, (Muñoz Cañavate & Chain Navarro, 2004b), and local administrative structures (Chain Navarro & Muñoz Cañavate, 2004). Recent issues of the IRIA and REINA reports document the evolution of these websites (MAP, 2008; Ministerio de la Presidencia, 2009). Also deserving mention are some very thorough and independent studies of the telecommunications sector that have been undertaken by different consulting firms and entrepreneurial foundations over the past years.

Despite the fact that other technological tools such as the mobile telephone or Digital Interactive Television are no doubt going to bear great relevance in the future, what is evident at present is that all administrative levels rely on the web as an essential resource for the growth of the e-administration. Internet was first implemented as a platform for the diffusion of information to the Spanish citizen in the mid-nineties, in a slow but steady process. The technology that had risen to power in the U.S. during the late seventies began to loom in Spain in the late eighties as a revolution withholding great repercussions for organizations in competition with other networks, over which it eventually prevailed (Bitnet, Fidonet, etc.); then it began to expand worldwide in the decade of the 1990s, under the auspices, so to speak, of the Gopher application in the first place, and of the World Wide Web afterwards.

3.2.1. General State Administration

Although the arrival of internet and its star application, the World Wide Web, was indeed a landmark for the process of interaction between bodies of the public sector and individual citizens, we must not forget that other technological innovations such as the videotex served for a time as the informational platform for providing services to users. Excepting France and its Minitel, the impact of such alternatives was negligent in Europe. Along with several other countries, Spain tried to set up services based on videotex. Table 3 shows the extent of Spain's experience with this service, which, here and elsewhere, was quickly made obsolete by the expansion of networks like the internet.

**Table 3**
Type of information providers in videotex.
Source: Ibertex Service Catalogs 1993 and 1995.

|  | 1993 | | 1994 | |
| --- | --- | --- | --- | --- |
|  | No | % | No | % |
| Public organisms | 96 | 32.88 | 193 | 34.59 |
| Private entities | 150 | 51.37 | 304 | 54.48 |
| Non-lucrative institutions | 46 | 15.75 | 61 | 10.93 |
| Total | 292 | 100 | 558 | 100 |

The IRA Report (MAP, 2008) identifies 701 units of the central administration as having a unit website, while in the previous edition there were just 375 units (MAP, 2006). However, the more recent report includes the headquarters of consulates, embassies, delegations of the Cervantes Institute and the CSIC Centers. It also indicates that in 31% of the cases, updating is done only when there are modifications of data; for 51% updates are daily, for 12% weekly, and for 6% monthly. Moreover, just 15% of the websites analyzed are safe sites –entailing the use of the https protocol– and only 13% of the websites of the central administration fulfill WAI norms of access for the handicapped. In 88% of cases there is no possibility of access by means other than the traditional computer, 11% are accessible by PDA and 5% by WAP.

The problems detected in the central administration are well illustrated in the INFOAGE report, put out by the Association of Computing Professionals of the State Administration of Spain (ASTIC). The report serves as a foundation for reflecting on the use of ICT in the Administration. One of the conclusions at which the ASTIC report arrived was that political leaders and policy makers have been slow to learn in the area of information technology, impeding greater speed in the process of change and modernization (INFOAGE, 2005).

There are indicators of 20 basic services. As a consequence of the comparative evaluative report of the European Commission within the Europe framework, 20 basic public services could be named in the administration, all of which are geared toward lending electronic services and distributing them among services to citizens and to businesses. Twelve of these services were directed to citizens, and eight to business. They were broken down into five types: taxes, registers, user facilities, individual or business revenue, and the processing of documents, permits, and licenses. The table below shows data from the European Union of 27 member states, and the countries that score above Spain in the 20 basic services. Whereas several countries have achieved 100% coverage, Spain is again in an intermediate position (Table 4).

**Table 4**
Internet availability of the 20 basic public services (%).
Source: Eurostat (2010). No data were gathered in 2005 or 2008.

|  | 2003 | 2004 | 2006 | 2007 | 2009 |
|---|---|---|---|---|---|
| EU (27) | – | – | – | 59 | 74 |
| Austria | 68 | 72 | 83 | 100 | 100 |
| Malta | – | 40 | 75 | 95 | 100 |
| Slovenia | – | 45 | 65 | 90 | 95 |
| Portugal | 37 | 40 | 60 | 90 | 100 |
| United Kingdom | 50 | 59 | 71 | 89 | 100 |
| Norway | 47 | 56 | 72 | 78 | 80 |
| Sweden | 67 | 74 | 74 | 75 | 95 |
| Germany | 40 | 47 | 47 | 74 | 74 |
| Spain | 50 | 55 | 55 | 70 | 80 |

*3.2.2. Administration of autonomous communities*

As in the rest of the administrations, the regional administration began its process of internet incorporation in the mid-nineties. Muñoz-Cañavate and Chain-Navarro (2004a,b) studied this process from 1997 to 2000, and deduced that the regional governments had established, with greater or lesser speed (depending on the region), systems of web information for the access of information on the internet. However, they also discovered that the evolution of the contents and services had not been so fast, and that the political
positions had become mired in a consideration of the WWW as a simple marketing possibility, without realizing that it was an outstanding tool of access to information for the citizen, which would speed up bureaucratic-administrative tasks, and that it could be a source of information about what was happening in each regional administration (Muñoz-Cañavate & Chain-Navarro, 2004a,b).

Evidently, 8 years later, the situation has changed enormously, as far as the e-administration of the regional governments is concerned; this leads us to comment on the Informe eEspaña 2007 produced by the Orange Foundation of France Telecom (eEspaña 2008, 2008).[22]

This study[23] followed and assessed the degree of development of the electronic administration of the 17 autonomous communities and two autonomous cites of Spain in terms of two indicators: the degree of total access of services, and the degree of sophistication of the service, which was judged in four stages. The first stage is the merely the supply of information, the final stage being the total lending of electronic services in question. The indicator used for evaluation was the online availability of 26 public services. Measuring the application of these services for the regions on the whole gave a figure of 67%; for services oriented toward citizens it was 69%; and the figure was somewhat lower for the 10 business-oriented services, at 64%. A service deemed to be in its fourth stage of evolution has reached the maximum level of maturity (eEspaña 2008, 2008). The following table reflects the availability of the respective services in each one of the autonomous communities of Spain (Table 5).

The results of the Fundación Orange study of 26 services reveal that online availability of services for complaints/suggestions prevails among the types of online services in existence (97%) and that the following four services score over 80%: university registration, public library consultation, ownership taxes, and documented judicial acts, and consultation regarding bidding for contracts. The services that meet the fourth stage of activity of full electronic service are complaints/suggestions in the first place, followed by university registration, payment of taxes, public libraries, and ownership taxes/judicial acts.

Finally deserving mention in this section is a study put out in Spain in 2008 by Biko2 consultants (Biko2, 2008) focusing on the websites of Spain's autonomous communities. This study of the usability of the websites of the autonomous communities underscores as a generalized error the use of their websites to announce political achievements of the regional governments. It also stresses that most of the autonomous communities have made a substantial effort to gain a position in the realm of citizen-oriented services, an aspect that can usually be corroborated with a glance at the homepage and the first pages of each section; when one delves deeper into the website, the improvements in usability, navigability and aesthetics decline very notably, to the point where the aesthetics may change, or information is suddenly structured according to the organic hierarchy of each administration.

Table 5
Availability of e-government services depending on the CCAA (%).
Source: España 2008. Fundación Orange and Capgemini.

| CCAA | Overall mean availability of the 26 services | Mean availability of the services to citizens (16 services) | Mean availability of the service to businesses (10 services) |
|---|---|---|---|
| Andalucía | 81 | 75 | 90 |
| Aragón | 63 | 69 | 53 |
| Asturias | 91 | 91 | 93 |
| Baleares | 59 | 58 | 60 |
| Canarias | 64 | 66 | 63 |
| Cantabria | 56 | 59 | 50 |
| Castilla la Mancha | 62 | 64 | 58 |
| Castilla y León | 69 | 72 | 65 |
| Cataluña | 74 | 77 | 70 |
| Comunidad Valenciana | 69 | 69 | 70 |
| Extremadura | 62 | 64 | 58 |
| Galicia | 77 | 81 | 70 |
| La Rioja | 67 | 72 | 60 |
| Madrid | 84 | 88 | 78 |
| Murcia | 72 | 75 | 68 |
| Navarra | 64 | 64 | 65 |
| País Vasco | 76 | 75 | 78 |
| Ceuta | 42 | 35 | 53 |
| Melilla | 41 | 52 | 28 |
| Media de España | 67 | 69 | 64 |

This study also signals the difficulty of getting through administrative tasks online, and the lack of services that allow paperwork to be done without any physical displacement to the administrative window.

### 3.2.3. Local Administration

Muñoz-Cañavate and Chain-Navarro began a series of studies in 1997 (2004) that analyzed the degree of internet implementation by the local administrations of Spain from 1997 to 2002. These papers, unlike other analyses, include the totality of the town halls across the country, and distinguished between official web pages of City Hall and other pages, applying a single methodology to define what exactly was meant by an institutional web page. It was often found, in the late nineties at least, that websites created by private citizens had been adopted by City/ Town Halls as property of their own (a practice virtually abolished since then). On a number of occasions the confused web phenomenon was observed, in which it was difficult to figure out who the web pertained to; or else the "off-and-on" website, which might disappear for some time only to reappear later on (Koelher, 2002).

Figs. 1 and 2 group town halls by their respective regions and demonstrate the development of Spain of Town Hall Government presence on the internet. Because of the great number of municipalities within Spain –over 8,000, nearly 7,000 having fewer than 5,000 inhabitants– and the fact that the smaller ones might have limited resources for creating informational policies on the internet, two graphs are presented. The first accounts for the totality of Spain's cities, towns and villages, and the second just for the City Halls of those municipalities with over 5000 inhabitants.

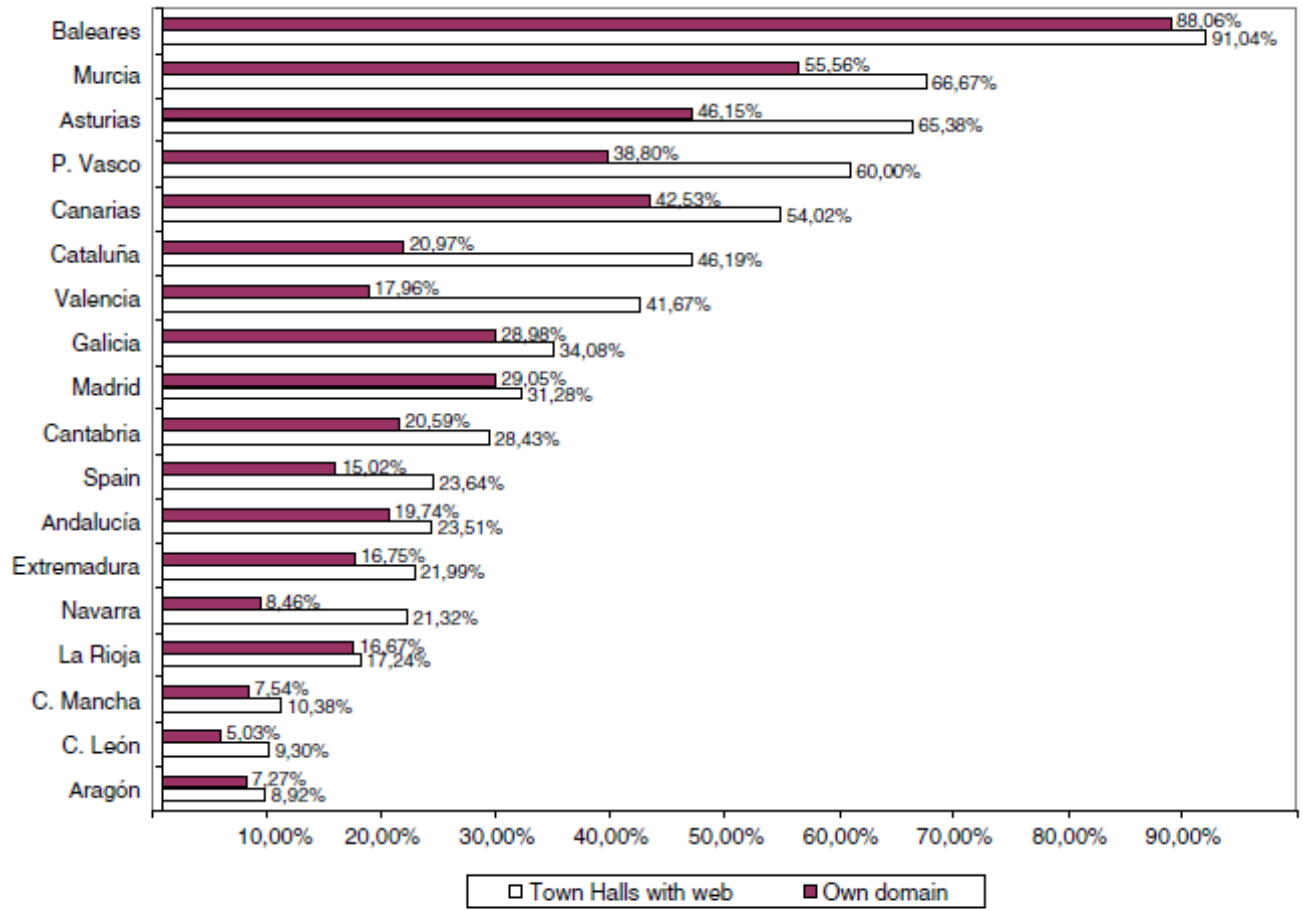

**Fig. 1.** Evolution of the City/Town Halls with respect to internet (1997–2002).

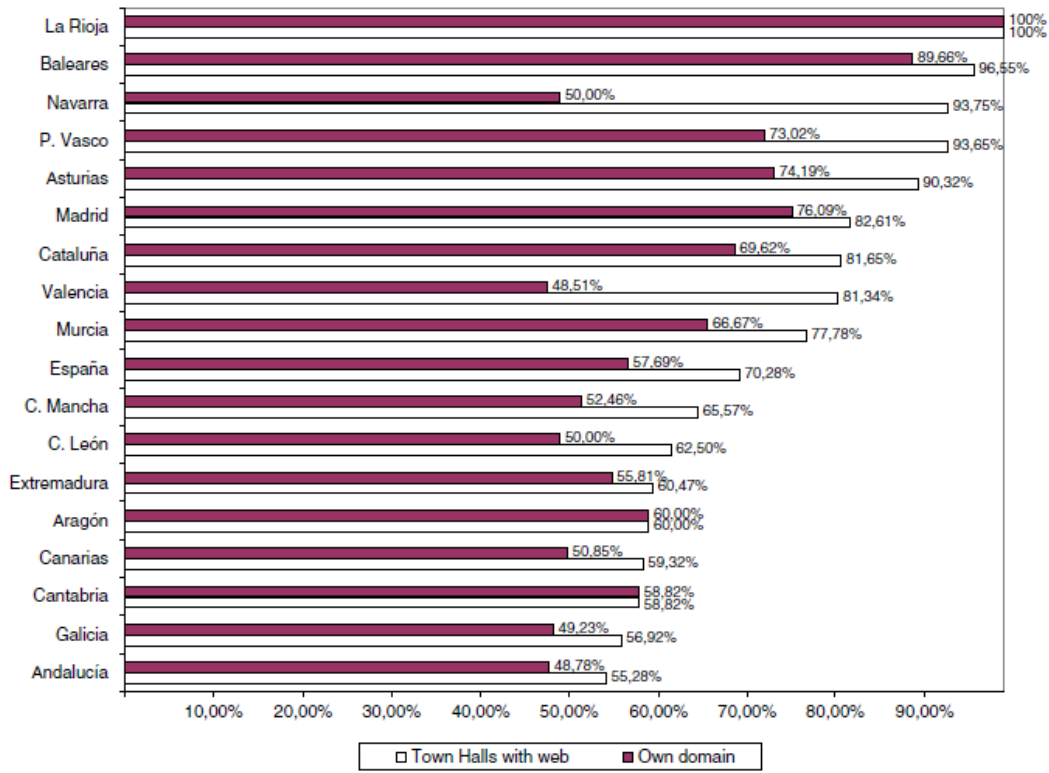

Fig. 2. Evolution of City/Town Halls of over 5001 inhabitants with respect to internet (1997–2002).

A comparison of two figures reveals another indicator, identified as a maturity index, signaling that the domains belong to the municipality itself, not a separate domain. In other words, a sub-domain may be registered under a domain of the first level (e. g. www.munimadrid.es) in contrast to the separate domains in which the institutional contents are found on other web pages (e. g. www.telefonica.es/munimadrid).

Fig. 1 demonstrates that 23.64% of the Town Halls across Spain can be found on the World Wide Web, and that 15.02% of them have domains of their own. As can be surmised from Fig. 2, with the larger municipalities these figures soar to 70.28% of City Halls with websites as of December 2002, 57.69% of them using domains of their own.

More recently, the IRIA report (published in 2008), presented the percentage of local entities with a website, something common to nearly 99% of the municipalities with over 10,000 inhabitants; and to 72% of the municipalities with populations between 1,000 and 10,000;

and to 50% of the villages of 500 to 1,000 inhabitants. However, this study does not define in its methodology what exactly is meant by a municipal website, something that was made explicit in the study referred to earlier (MAP, 2006).

The IRIA 2008 report found that 50% of the municipalities with over 100,000 inhabitants allow the use of credit card for payment via the web page; the smaller the towns, the lower this percentage. It also reported that the entities who comply with norms for web access is 100% in the large municipalities, yet drops to 50% for the smallest ones. Moreover, 100% of the large municipalities offer some service(s) completely online.

In order to explore the services offered by Spain's Town Halls, we must cite other studies. Software AG, for instance, put out a study titled "Analysis of the Development of electronic Government in Spain" (*Análisis del Desarrollo del Gobierno electrónico Municipal en España*) (Software, 2005) in which 91 Spanish cities were evaluated (provincial capitals and cities with over 75,000 inhabitants). One conclusion arrived at through the Software AG study was that many cities try to offer internet services, but that they do so in a fairly impractical manner. In addition, this study finds that many populations require that users finalize a service begun online either on the telephone or in person at the town hall. The analysis also reflects worrisome evidence that the information in question is not up-to-date. It concludes, moreover, that the persons in charge of city government should make a greater effort to implement digital certification and payment over the web (Esteves, 2005).

Meanwhile, the report *eEspaña* 2006 by the Orange Foundation analyzed the websites of 390 town and city halls regarding five areas: security, information, navigability and design, contents, and services and participation (eEspaña 2006, 2006). This study found very limited security, with only 11% having a server perceived as "safe." Noteworthy findings included the scarce number of websites offering personalized information—data such as income declaration, administrative records, the payment of municipal fees, and the census. Navigation and design had improved with respect to the previous year, as had web accessibility for the handicapped. Some 30% of these portals featured multimedia materials (videos, virtual visits, recordings of plenary sessions). The most usual contents refer to the composition of the organism, public employment, as well as information about plenary sessions, budget, and ordinances. The study also showed that most city halls have very limited electronic services (the most frequently offered service being online economic transactions, for 21.8% of the town halls). Finally, even though the study identified that 60% of the municipalities offered a suggestion box for citizens, it concluded that citizen participation was lower than might be anticipated.

The study by Chain Navarro et al. (2008a) mentioned above concludes that, at the municipal level, Spain is still in the early stages of informational offers. Granted, some local Administrations are advancing quicker than others; but in general, e-government lies on a distant horizon.

## 4. Levels of access to citizens and enterprise: internet and electronic administration

This section explores data on internet access from homes and citizens, as well as from businesses and employees throughout Spain; these figures will be compared with those of the European Union. In addition, we can look at the percentages of citizens and businesses that relate with the administration, and for what purposes. In all cases we present the official data of Eurostat.

Tables 6 and 7 reflect the extent of introduction of internet in European homes, and internet access on the part of European citizens aged 16 to 74 (having used internet at least once a week over the 3 months previous to the study). The Spanish data reveal less growth than among Europe on the whole (EU 15 and EU 27) in both cases.

**Table 6**
Percentage of homes with internet access[a].
Source: Eurostat (2010).

|         | 2003 | 2004 | 2005 | 2006 | 2007 | 2008 | 2009 |
|---------|------|------|------|------|------|------|------|
| Spain   | 28   | 34   | 36   | 39   | 45   | 51   | 54   |
| EU (15) | 43   | 46   | 53   | 54   | 59   | 64   | 68   |
| EU (27) | –    | 40   | 48   | 49   | 54   | 60   | 65   |

[a] The data presented correspond to the last ones published each year.

**Table 7**
Percentage of citizens (aged 16 to 74) with internet access.
Source: Eurostat (2010).

|           | 2003 | 2004 | 2005 | 2006 | 2007 | 2008 | 2009 |
|-----------|------|------|------|------|------|------|------|
| Spain[a]  | 29   | 31   | 35   | 39   | 44   | 49   | 54   |
| EU (15)   | 38   | 41   | 46   | 49   | 55   | 60   | 64   |
| EU (27)   | –    | 36   | 43   | 45   | 51   | 56   | 60   |

[a] Another survey, by the firm Sofres para Red.es regarding the population of Spain over age 15 connected to internet, gave findings that were higher, obviously. Internet users numbered: 42.2% in 2004, 46.6% in 2005, 50% in 2006 and 52.4% in the first trimester of 2007.

In Table 6 we corroborate that mean use across the European 15 member states[25] is 14 points higher than the use in Spanish homes connected to the internet[26] (45% as opposed to 59%); the same lag is seen for the percentage of citizens with internet access. However, in the business realm, as we see in Tables 8 and 9, the figures for firms connected to internet and for employees with access to the network from their workplace are more or less on par with the rest of the European countries.

**Table 8**
Percentage of businesses with internet access.
Source: Eurostat (2010).

|         | 2003 | 2004 | 2005 | 2006 | 2007 | 2008 | 2009 |
|---------|------|------|------|------|------|------|------|
| Spain   | 82   | 87   | 90   | 93   | 94   | 95   | 96   |
| EU (15) | 85   | 91   | 92   | 94   | 95   | 95   | 96   |
| EU (27) | –    | 88   | 91   | 92   | 93   | 93   | 94   |

**Table 9**
Percentage of employees with internet access from the workplace.
Source: Eurostat (2010).

|         | 2003 | 2004 | 2005 | 2006 | 2007 | 2008 | 2009 |
|---------|------|------|------|------|------|------|------|
| Spain   | 16   | 18   | 20   | 22   | 23   | 25   | 26   |
| EU (15) | 19   | 21   | 23   | 24   | 27   | 29   | 30   |
| EU (27) | –    | 18   | 21   | 22   | 25   | 26   | 27   |

Tables 10 and 11, in turn, demonstrate the percentages of citizens and businesses that, with regular access to internet, have actually used the network to interact with public authorities. Again, the situation of Spain is well behind the mean of other European countries.

Finally, Tables 12 and 13 give the Eurostat data from two different studies where citizens and businesses were asked about their use of internet in dealing with the Public Administrations. In both cases, it was reported that the internet was used for obtaining information, downloading applications and forms, and sending in forms. Spain again shows lower percentages in comparison with the average for EU 15 and EU 27 under all the indicators.

**Table 10**
Percentage of citizens (aged 16 to 74) who have used internet to interact with public authorities.
Source: Eurostat (2010). Only existent data are given.

|         | 2005 | 2006 | 2007 | 2008 | 2009 |
|---------|------|------|------|------|------|
| Spain   | –    | 25   | 26   | 29   | 30   |
| EU (15) | 26   | –    | 34   | 32   | 33   |
| EU (27) | 23   | 24   | 30   | 28   | 30   |

**Table 11**
Percentage of businesses that have used internet in their relations with public authorities.
Source: Eurostat (2010). Only existent data are given.

|         | 2003 | 2004 | 2005 | 2006 | 2007 | 2008 | 2009 |
|---------|------|------|------|------|------|------|------|
| Spain   | 44   | 50   | 55   | 58   | 58   | 64   | 65   |
| UE (15) | –    | 50   | 56   | 64   | 66   | 70   | 74   |
| UE (27) | –    | 51   | 57   | 63   | 65   | 68   | 71   |

**Table 12**
Share of individuals using the internet for interacting with public authorities.
Source: Eurostat (2009).

|         | For obtaining information from public authorities | For downloading official forms | For sending filled forms |
|---------|---------------------------------------------------|-------------------------------|--------------------------|
| Spain   | 28.6                                              | 15.7                          | 8.5                      |
| EU (15) | 30.6                                              | 19.3                          | 14.4                     |
| EU (27) | 27.5                                              | 17.5                          | 12.7                     |

**Table 13**
Share of enterprises using the internet for interacting with public authorities.
Source: Eurostat (2009).

|         | For obtaining information | For downloading official forms | For returning filled in forms |
|---------|---------------------------|-------------------------------|-------------------------------|
| Spain   | 60                        | 59                            | 46                            |
| EU (15) | 66                        | 66                            | 56                            |
| EU (27) | 65                        | 64                            | 55                            |

# 5. Running through the processes of modernization in Spain

This fifth section looks at the political initiatives put into effect by diverse means and that have proven to be the most relevant within the State Administration. As we mentioned early on in this paper, both the decentralization of Spain's political-administrative system and the fact that the regional governments have full political capacity to define their proposals with regard to citizens and business (as does the local Administration) make the study of Spain's implementation of e-administration very complex. For this reason, from this point forward, we focus our discussion on the initiatives of the central government. We do this however with the conviction that many aspects of regional and local administration are propelled by the central government, either because the projects of the central government were carried out in collaboration with the other Administrations, or because initiatives designed in Madrid were imitated by regional and local administrations.

It is remarkable how many projects for modernizing were not avidly accepted by the body of citizens, or whose existence was never even known. The latest survey of quality in public services of the *Centro de Investigaciones Sociológicas*, or CIS –an official organism dedicated to carrying out sociological studies– dating from October 2006, reports that 78% of citizens had never heard of the governmental measures undertaken to enhance the quality of public services (CIS, 2006b).

*5.1. I Plan for the modernization and improvement of the State Administration in 1992*

The arrival of Democracy in Spain during the second half of the 1970s set into motion a process of legal reforms intended to install an administrative system that would accommodate this new situation while also lending it greater efficiency, so that the initial focus of attention was on the internal elements of organization, which would extend over time, throughout the 1980s.

From that point on, Spain saw a series of modernizing processes established with the clear underlying notion of making the citizen the axis of innovative efforts. To this end, there is no questioning the importance of the OECD document "*Administration as Service*," which was published in Spanish in 1988 and served as a catalyst of reflections gestating within the Administration, targeting a more pragmatic public organization.

This fertile terrain gives rise, in 1992, to the "First Plan for the Modernization and Improvement of the State Administration" (*I Plan de Modernización y Mejora de la Administración del Estado*, 1992), which allows the central administration of Spain on the whole to initiate a process of overall change. Its foundations lie in a study by MAP entitled "Reflections about the Modernization of the Administration of the State," which came out in 1989 and underlined the need to redefine our Public Administration.

The first phase (from 1992 to 1993) entailed the design of three strategic lines that embraced a total of 204 programs:

- Improvement of information and the communication of information to citizens and businesses;
- Improved quality of services; and
- Improved internal management (administrative simplification, formation, human resources, control and evaluation, and cooperation).

The first of these lines took in 41 projects covering four basic areas: bringing the Administration closer to the citizen; creating new information offices; using telematic means for accessing information; and improving instruments of information and diffusion for the promotion of economic, cultural, and social activity. In addition to the aforementioned strategic lines of action, the second phase would involve a fourth line of action, the "reduction of costs and increase in productivity," and would distribute the 41 projects across all four lines of action. Table 14 gives a sample of the information services produced by the AGE; no services involving internet were available at the time (MAP, 1995).

During the second half of the nineties, various initiatives were aimed at improving quality and the coordination among Administrations also became more effective than in previous years. In terms of the repercussions for citizens in general, the most important were the "One-stop Service" (*Ventanilla Única*) and "Business One-stop" (*Ventanilla Única Empresarial*) projects, which affected several administrations; the introduction of the Letters of Service and Quality; and the "white books" for the improvement of public services. The One-stop Service and Business One-stop projects –both of which were implanted by the central government in 1996 to ensure the progressive utilization of a intercommunicative system of records between the State administration and the regional and local Administrations– no doubt served as models of cooperation among administrations. However, the central government itself understood that in 2007 their scope was insufficient: with the One-stop Service, the citizen could only present applications or documents to other administrations, and the Business One-stop was a project that did not apply to all entrepreneurs of any locality. In this way, the agreements that had given rise to the One-stop Projects were replaced by an Agreement of the Council of Ministers, dated June 15, 2005, which created a network of integrated Offices of Attention to the Citizen in collaboration with regional governments and local administrative bodies (Tables 15–17).

**Table 14**
Online electronic information services, CdRom, diskette, magnetic tape, videotext, and audiotex.
Source: Directory of electronic information services of the AGE (1995).

| Ministries and other organisms | Services produced | |
| --- | --- | --- |
| | Number | % of total |
| Foreign Affairs | 6 | 1.4 |
| Justice and Interior | 17 | 4.0 |
| Economy and Treasury | 73 | 17.0 |
| Public Works, Transportation, and Environment | 42 | 9.8 |
| Education and Science | 31 | 7.2 |
| Labor and Social Security | 41 | 9.6 |
| Industry and Energy | 23 | 5.4 |
| Agriculture, Fishing, and Food | 7 | 1.6 |
| Public Administrations | 47 | 11.0 |
| Culture | 26 | 6.1 |
| Presidency | 38 | 8.9 |
| Health and Consumers | 31 | 7.2 |
| Social Affairs | 23 | 5.4 |
| Commerce and Tourism | 8 | 1.9 |
| Others | 16 | 3.7 |

**Table 15**
Catalog of the services of Information Society.
Source: Info XXI.

| | |
|---|---|
| Services of information to the exterior | 103 |
| Services of internal information | 31 |
| Service of communication with the exterior | 10 |
| Services on internal communication | 25 |
| Service of teleadministration | 27 |
| Teleeducation, teleformation, telework, and telemedicine | 19 |
| Foment of technological development and the use of new technologies for businesses | 44 |
| Legal and normative framework | 44 |
| Improvement of internal management | 16 |
| Development of new information society services and promotion of their use by citizens and businesses. | 28 |

**Table 16**
Measures contemplated in the España.es plan for e-administration development.
Source: España.es.

| | |
|---|---|
| Measures for facilitating public access for users | Electronic Identity Document |
| | Points of free public access to internet in the offices of registration and public attention of the AGE, and in the offices featuring Business One-stop and in those town halls belonging to the One-stop Service convention |
| Measures for promoting the development of user services | Foment the development of basic electronic public services under eEurope 2005 |
| | Letters of electronic services |
| | Permanent development of the Citizen's Page |
| | Utilization of the co-official languages and other languages on the AGE web page |
| | Accessibility to the web pages of the General State Administration |
| | Forms on internet and creation of telematic registers |
| | Secure telematic notifications |
| | Payment over internet |
| Measures for facilitating the interchange of information among the Public Administrations | Substitution of paper certificates, interchange of telematic certificates and transmission of data |
| | Page of services of the AGE for the entities that make up the Local Administration |
| | Migration of internal communications of professional associations regarding ICT to telematic channels |
| Measures for supporting the internal reorganization in the Public Administrations | Foment the Public Employment Page |
| | Reform of the Upper Council of Automation (Consejo Superior de Informática) and for the promotion of e-administration |
| | Coordination with the territorial administrations for the promotion of e-administration |
| | Revision and analysis of administrative procedures of the AGE to allow for their telematic application |
| | Service of technical support to the Ministerial Departments |
| | Service of archives for electronic documentation |

**Table 17**
Services for the citizen under Spain's *Seguridad Social*.
Source: TGSS (General Treasury of Social Security).

| | |
|---|---|
| Services without digital certificate | Report of work history |
| | Report of basis of payments |
| | Consultation of status and request for reports |
| | Reports on basis and quotas paid in fiscal period specified |
| | Auto-calculation of special contractual terms |
| Services with digital certificate | Report of work history |
| | Application for rectification |
| | Report of payments |
| | Duplicate of document of affiliation |
| | Request for change in payment base (free-lance or autonomous workers) |
| | Situation of payment for independent work |

*5.2. Plan Info XXI: Information Society for all*

The year 1996 was witness to the baptisms of fire of the VI Legislature. When the internet first came into being –in an international context where initiatives and programs involving the information society proliferated– there arose in Spain the "Plan Info XXI" (subtitled "Information Society for All"), which was approved by the Council of Ministers on December 23, 1999. Some months before, in July of 1999, Spain's Interministerial Commission on the Society of Information and New Technologies had been created, its main objective being to coordinate the initiatives that, until then, had been promoted within various bodies of the AGE. This commission, together with the *Consejo Superior de Informática* of the MAP and the *Consejo Asesor de Telecomunicaciones* of the Ministry of Public Works, put together a catalog of all the Information Society projects under the umbrella of the AGE, which are listed below (Muñoz Cañavate, 2003).

Plan Info XXI did not harvest spectacular results by any account. The criticism that hurled from the media and political parties in opposition to the Plan was accentuated by a report that had been requested by the government itself. In evaluating the Plan Info XXI, a Special Commission for the Study of Development of Information (*Comisión Especial de Estudio para el Desarrollo de la Sociedad de la Información*) had been constituted in November 2002. Otherwise known as the "Soto

Commission," after the name of the person presiding it (Juan Soto-Serrano), it brought to light a number of obstacles for IS development within Spain:

(a) Insufficient leadership, disembarking in initiatives that did not attain their objectives;
(b) inefficient management, a lack of coordination, and duplicity of efforts;
(c) a lack of awareness on the part of citizens of the advantages of ICT;
(d) high complexity associated with the incorporation of ICTs into the processes of public and private organizations.

5.3. *España.es program*

The recommendations of the "Soto Commission" gave way to approval of a program known as "España.es," whose measures for encouraging electronic administration were structured along six major lines, three of them vertical (e-administration, education, and small/medium-sized businesses) and three horizontal (accessibility and formation, digital contents, and communication).

5.4. *Avanza*

Nonetheless, *España.es* was not the final effort. The arrival of the VIII Legislature saw the design of a new plan: "Ingenio 2010," which was to entail a clear orientation toward research. Indeed, it stood as a joint strategy of promoting Research, Development and Innovation (*Investigación + Desarrollo + innovación*). This new plan was devised with three key programs at its core. Namely, they were "Cenit" (*Consorcios Estratégicos Nacionales de Investigación Técnica*), aimed to increase public and private cooperation in I + D + i; "Consolider," with the goal of enhancing research excellence; and "Avanza," dedicated strictly to the Information Society.

Avanza has been structured, to date, along two plans, Avanza (from 2005 to 2008) and Avanza 2 (since 2009). In 2005, the design of Avanza was the indirect result of the diagnosis of Spain's situation in the wake of the elections of 2004, with a new governing party far removed from the party and ideology that had governed Spain over the previous 8 years. The Socialist Party, PSOE, had won the general elections after Conservative PP, or Popular Party, had served two terms in office. The conservatives, who had promoted plans InfoXXI and España.es, were in clearly precarious positions.

Avanza was devised along these five lines of action:

> 1. Homes and the inclusion of citizens—to increase the use of ICT in homes, and citizens' everyday participation in public life.
> 2. Competition and innovation—to make ICT stronger in the private sector.
> 3. Education in the Digital Era—to incorporate ICT in the educational process.
> 4. Digital Public Services—to improve Public Administration Services.
> 5. Digital Context—to expand use of the wide band.

The fourth line, "Digital Public Services," is concerned with e-administration in different types of services: State Administration and Local Administration, Health Services, Education and Justice. The State Administration is in charge of 33 very diverse projects, ranging anywhere from the introduction of services to the installation of infrastructures. Included among the e-services offered by the State Administration are: telematic notifications that are safe for citizens and businesses, direct payment by telematic means, elimination of the need to present paper documents, electronic contracting, and unified 060 information services. Included among the infrastructures of e-administration under the State Administration, is the Sistema de Aplicaciones y Redes para la Administración Pública (Applications and Networks System for Public Administration) Project (SARA), to be described below.

The projects dedicated to the local Administration (Avanza Local) revolve around three lines of development:

(a) Solutions for local entities whose objective it is to help administrations of the local realm in improving their *back-office* and *front-office* processes for citizens and businesses.
(b) Digital Cities, which include the program begun under the conservative government of 2003 that was dedicated to implementing Information Society in the local setting, and included very diverse spheres of action: teleadministration, teleemployment, telemedicine, formation, diffusion and awareness, electronic commerce/business, culture, tourism and free time, and applications for social groups with special needs.
(c) The Digital Town Hall program, begun in 2006 in substitution of Digital Cities, with objectives similar to those of the former.

Finally, Avanza specifies three further areas of action: education toward technological literacy, or the implementation of wireless networks; e-administration developments in health care that includes a coordinated project involving the central government and regional governments, to achieve full computerization of medical assistance and the extension of new health-related technologies (making appointments via the internet, consulting health-related information, telediagnosis, telemedicine, etc.); and e-administration implementation in the area of justice, a domain traditionally forgotten in Spain when it comes to ICT. This final aim was to culminate the internal process of modernization of the plan initiated in 2002 by the previous government, and to computerize the civil registers and justices of the peace.

Plan Avanza 2 aims to foment the demand for ICT and promote the consolidation of an ICT industry within Spain specializing in strategic sectors, with the small/medium business foremost in mind. Avanza 2 offers Digital Public Services that will improve the quality of those web services already extended by the Public Administration, with an emphasis on support of Local Entities and the use of an electronic ID card.

## 5.5. The emblematic projects of the AGE

In the complex structure of the central administration, then, all of the ministries have provided services that can be accessed over the web, and have undertaken with greater or lesser success reforms to computerize the internal management of procedures, making them readily available to users, through systems of electronic signature. In State government, it is common to refer to various emblematic projects stemming from recent legislatures and from governments of distinct political natures; these include the *Agencia Estatal de Administración Tributaria* (State Agency for Income Administration), *Tesorería General de la Seguridad Social* (General Treasury of Social Security), the CERES project of the *Fábrica Nacional de Moneda y Timbre* (National Factory of Money and Stamps), *Oficina Virtual del Catastro* (Virtual Property Register), the "red 060.es" network, and the aforementioned SARA platform.

*5.5.1. Income taxes: Agencia Estatal de Administración Tributaria*

The *Agencia Tributaria* first set sail on the internet in 1996, though one precedent of it can also be found in the videotex system. Its evolution has been noteworthy, to the point where, at present, it is considered to have reached a state of maturity that permits the full-fledged processing of income tax declarations and returns via internet.

*5.5.2. Treasury and Social Security: Tesorería General de la Seguridad Social*

Spain's Social Security and public health system (*Seguridad Social*) offers diverse services to the citizen, who does not need to be physically present at their offices. In some cases the user and Administration interact without a digital certificate, and in other cases it is indeed used. Services involving certification require an electronic national identity document (DNI), or a different certificate that can be expedited in Spain by the National Factory of Money and Stamps.

At the same time, they have a system for the electronic issue of documents, known as RED, that can move documents among enterprises, professionals, and authorized representatives of the General Treasury of Social Security, and which uses internet as its medium. The processing stages that can be carried out include medical leaves, changes in the level of earnings and taxes, contract modifications, reports on the status of payments, or consultation as to the situation of business affiliates.

*5.5.3. CERES (Certificación Española)*

This project materializes within the framework of e-administration projects requiring authentication of user identity when transactions involve the administration. With its onset in 1996, and full validity in 1999, the project is carried out under the auspices of the *Fábrica Nacional de Moneda y Timbre*, and in essence it establishes a public entity of certification that seeks to ensure confidentiality in communications among citizens, businesses, and the public administrations. The identity of users is stored in an intelligent card that is only accessible for the user by means of a personal identification number, although the cryptographic profile is also stored, and if a card is not used, one can gain access to the file using a PIN.

*5.5.4. Property Register: Oficina Virtual del Catastro*

Available since 2002, this is another of the predominant projects of State Administration. It makes it possible to process property taxes with no need for physical presence, by means of certificates expedited by Spain's *Fábrica Nacional de Moneda y Timbre*.

Property taxes are of great administrative importance given that the statistical census and register of all rustic and urban properties requires absolute authentication of all data. This context takes in different administrators of information: the citizens themselves, who are obliged to present declarations of the modifications and transmission of estates of which they are owners; the local administration, which supplies information about urban changes; the notaries and property registrars, who offer information about alterations appearing in public notices and the new inscriptions in the property register; and other administrations that may provide information about the territory—for instance, the building of highways or major public works would call for a modification in the ownership of expropriated terrain (Alonso Peña, Fernández Gómez, & Yánez Morante, 2007).

The service permits:

- Consultation of the data regarding real estate or property register on the part of the owner and the Public Administrations.
- The interchange of files with relevant data among the agents of the
   Property Register and other Administrations (for example, notaries).
- Visualization of property maps.
- Requests for property-related data.

### 5.5.5. *Attention to the citizen: 060 offices*

Attempts by the central Administration of Spain to gather onto a single website all of its user services can be traced back to September 2001, when the *Ministerio de Administraciones Públicas* (MAP) created the website *administracion.es* to channel information from the Public Administration to citizens and businesses (Muñoz Cañavate, 2003).

Later, the new government of 2004 transformed that website into the so-called "servicios 060." Thus, 060 becomes a network encompassing the offices for attention to citizens and businesses, through two means: a telephone number (060) that serves over one thousand numbers of different organisms, and a website *060.es*, which replaces the previous one (administracion.es) and allows information to be accessed in three separate ways: by topic, by user profile (e. g. youth, seniors, immigrants, tourists), or attending to vital facts (e. g. marriage, retirement, search for housing). However, different blogs have expressed criticism of the lack of administrative steps in processing and, above all, the strange sensation of jumping from an initial page with one design to subsequent pages of the Administration with designs and structures that are different.

Some also believe that the era of websites such as *administracion.es* or *060.es* has passed, and that the system of information retrieval should now look toward models like that offered by Google to U.S. citizens ([www.google.com/ig/nsgov](www.google.com/ig/nsgov)), where the user can write in the language that he or she prefers.

Precisely at the beginning of 2008, in Spain we were surprised by news that the government, through the Ministry of Public Administrations, had contracted Google technology to centralize all the searches for information of all ministries.

### 5.5.6. Infrastructures: SARA

Infrastructures configure another cornerstone of e-administrative development. In this context we find SARA, a system of applications and networks that allows all interested public administrations to be connected by an administrative extranet and a service-based architecture. Although Spain's administration is tremendously complex, by connecting different administrations in this manner, the procedures for citizens are simplified. SARA makes it possible to bypass the need to photocopy one's National Identity Document (DNI) or paper voting registers or census sheets; it also eliminates the requirement of presenting certificates on paper form, which in turn facilitates not only citizen interaction with the administration, but also the work of the Administrations themselves. The load of paperwork decreases, so that the relations with the citizen are simplified and the attention to and the conditions of lending public service are improved, with reduced waiting lines for the public, and a significantly reduced volume of paper documents—reduced to only those that are entirely necessary.

## 6. Public employees

This section draws to a close our general overview with reference now to the role of public employees. Over the past two decades, the management of human resources has acquired a predominating role in the management of organizations, yet this is also true within the Public Administration. Allison established a model of administrative functions based on three axes: strategy, the management of internal elements, and the management of external elements.

Such internal elements would include human resources and staff management as keys to the success of any project in the public sector. It is understood that public employees constitute an essential catalyst for administrative transformation. No e-administration project is possible without the support and effort of the personnel of which these administrative bodies are comprised. And though the public employee stereotype has not been particularly positive in the recent history of Spain, we may also affirm that advancement in policies involving the staffing of these administrations has progressively changed work habits and manners, and consequently, the citizens' perception thereof.

Still, the aforementioned survey in October 2006 by the CIS, to measure the quality of public services, made it known that 54% of those surveyed had a neutral or negative impression of the workings of the public administration, and only 34% held a positive impression. No doubt, the public employees are partly responsible for these findings (CIS, 2006b).

Formation and training are considered essential, for which reason the Law of Electronic Access of Citizens to the Public Administrations states that public employees of the General Administration of the State will receive specific formation to guarantee updated knowledge of the conditions of security in the utilization of electronic means in administrative activity, as well as the protection of data of a personal character, the respect of intellectual and industrial property, and the management of information (BOE, 2007a).

The so-called "Plan Concilia," an integral plan intended to reconcile personal life and working life in the state administration, includes among its measures the application of telework techniques in the public sector. It defines telework as "a form of working with information, by means of telematic tools and without dependence on a concrete space. It tries to take advantage of these technologies fundamentally with respect to the flexibility that they afford in terms of space and time".

Yet not all public employees are equally involved in the e-administration. One significant datum is the percentage of public employees with email: whereas in the state administration 54% of employees overall have electronic addresses of their own (although 100% of the strictly administrative employees do), in regional administrations the figure is 69%, and at the local administrative level just 46% (Fundación Telefonica, 2007).

In Spain, the CIS polled public employees in 2006, and their findings are given below as the most recent "barometer" of opinion, from a sample of 1464 employees (CIS, 2006a). We chose three of the questions on that survey to have a closer look. The results, shown in Tables 18–20, clearly reflect how public employees have naturally adapted to the everyday application of new technologies, and hold that it has meant better service. Noteworthy is the percentage of responses that manifest a need to adapt the administrative processes to new technologies, the infrautilization of these in many cases, and the inability of many employees to adapt to the use of the new technologies.

**Table 18**
Regarding the use of new technologies in the Administration, there are very diverse opinions. Do you largely agree or disagree with the following notions?
Source: CIS. Estudio 2064 Funcionarios Públicos, 2006.

|  | Largely agree | Largely disagree | Do not know | No answer |
|---|---|---|---|---|
| The use of new technologies has proven to be hardly efficient at all, because the administrative processes have not adapted to their use | 25.2 | 72.3 | 1.2 | 1.2 |
| Thanks to the use of new technologies, the workings of the Administration have improved | 92.9 | 5.8 | 0.9 | 0.4 |
| In general, in the Administration one sees an infra-utilization of the facilities/equipment of the new technologies | 48.7 | 46.6 | 3.5 | 1.2 |
| Many of the present public employees are incapable of adapting to the use of the new technologies | 35.9 | 60.5 | 2.8 | 0.8 |

**Table 19**
To what extent (great, fair, some, little or none) do you believe that a greater use of these new technologies will serve to improve your production overall?
Source: CIS. Estudio 2064 Funcionarios Públicos, 2006.

| | |
|---|---|
| Much | 40.6 |
| Quite a bit | 47.4 |
| Little | 7.9 |
| Hardly at all | 3.1 |
| Do not know | 0.3 |
| No answer | 0.6 |

**Table 20**
Do you agree (strongly, somewhat, scarcely or not at all) that the adaptation to the use of new technologies is a basic criterion for the reorganization of the administration in coming years?
Source: CIS. Estudio 2064 Funcionarios Públicos, 2006.

| | |
|---|---|
| Strongly agree | 37.5 |
| Somewhat agree | 48.9 |
| Neither agree nor disagree | 3.9 |
| Scarcely agree | 7.1 |
| Do not agree at all | 1 |
| Do not know | 1.2 |
| Do not answer | 0.5 |

## 7. Conclusions

Having described the horizon of legislative and political initiatives, it is time to sum up the situation of Spain as an intermediate one. Despite firm investment and diverse strategic planning, with varying degrees of success, Spain does not manage to take off or stand out in any of the areas of administration, as corroborated by the comparative statistics of Eurostat. Nonetheless, the effort towards legislative development in the right direction must be acknowledged.

While some specific initiatives such as those described in this article – most notably Spain's *Agencia Tributaria and Seguridad Social*– have completed programs of fully electronic services, the advances are generally slow in coming. And there has been much debate about the reasons why. One possibility is that the major administrative moves take place transitions from one party to another in Government; another possibility is that even when persons within the same political party are renewed, earlier policy tends to be regarded as failed policy, or else the achievements of others are swept under the rug, in either case ignoring the fact that accumulated experience is always fruitful.

Nor is the dispersion of efforts resulting from administrative decentralization a good thing (though it could be otherwise). We believe that decentralization should be compensated by a better coordination of efforts, so as to avoid the development of those "islands of progress" that have characterized the Spanish panorama up to the present.

In our opinion, an Administration as complex as that of Spain must have mechanisms for regular assessment of the steps taken by the different governments, just as we do to measure the number of users or businesses with access to internet. Moreover, in order for the results of a study to generate sufficient credibility, it must be undertaken by an independent entity.

Various approaches and routes to progress have been taken, even though the end results have often been the same. Unfortunately, this means that the collective experience has been ignored. In light of this reflection, we can appreciate the beneficial agreement reached by the Orange Foundation of France Telecom and Capgemini, who in recent years have gained increasing credibility with their periodical reports on the level and situation of e-administration services in different countries. At the same time, we understand that certain data from reports such as IRIA, REINA or CAE can be effectively supplied by their own administrations (for instance, the data on investment); yet other indicators, such as those needed to evaluate the quality of services provided, should be carried out by external entities, whose objectivity will allow all political factions to have reliable information at their disposal.

**Notes**

1 The figures of the Instituto Nacional de Estadística speak for themselves: in 1998 the official population of Spain was 39,852,651 inhabitants, whereas in 2009 the figure had risen to 46,745,807.

2 The political underpinnings of the results obviously allow reality to be masked when the affected parties themselves are the ones who evaluate their services, instead of an independent entity.

3 REC 2004 15 "Electronic Governance".

4 Communique of the Commission, September 26, 2003, to the Council, the European Parliament, the European Economic and Social Committee and the Committee of the Regions. COM (2003) 567 final.

5 The project was promoted by Spain's *Ministerio de Industria*, Red.es, and the Spanish Federation of Municipalities and Provinces, which takes in 85% of the local bodies of Spain. More specifically, the project consists of the digitalization of all the information relative to urban planning, the implantation of a system of geographic information, and its publication on Internet.

6 The study "La Sociedad de la Información en España 2004" gathers some studies in different parts of the world that reveal the savings in cost through electronic administration. One study done in Canada focuses on the savings in a service transaction: in person, it comes to 44 Canadian dollars, by mail 38 dollars, by telephone 8 dollars, and online, less than one dollar. It also states that Ireland´s online income tax service means savings of 33% as compared to manual processing, thanks to the online presentation of the income tax declarations, and that 600,000 euros have been saved by eliminating paper forms. Finally, it mentions the case of Spain's Agencia Tributaria, which, thanks to computerized processes, spends only 68 cents for every
100 euros that it gets back, one of the lowest figures among the OCDE. *La Sociedad de la Información en España 2004*. Madrid: Telefonica, 2004, p. 179.

7 However, some authors believe this article lacks clarity, understanding that this right depends on the media (Cornella, 1998).

8 We refer to the *Jornadas de Informática de la Administración Local* (JIAL), which began in 1979, and the *Jornadas sobre Tecnologías de la Información para la Modernización de las Administraciones Públicas* (TECNIMAP), appearing in 1989. It was also in 1989 when the Ministerio de Administraciones Públicas put out the document "*Reflexiones para la Modernización de la Administración del Estado*".

9 This law was partly modified in 1999 and affected in 2001 by Spanish Law 24/2001, of 27 December, of Fiscal, Administrative and Social Measures (*Medidas Fiscales, Administrativas y del Orden Social*).

10 For this purpose, any citizen or business must have a Unique Electronic Address, subscribe to the procedures chosen, and receive notifications by e-mail. http://notificaciones.administracion.es/.

11 The updated figures, as of September 2007, list 15 servers of certification having effected the communication required by Spanish Law 59/2003 of electronic signature. One such server is the National Mint or *Fábrica Nacional de Moneda y Timbre*, which can give an electronic ID card (DNI). Retrieved April 10, 2008 from http://www.mityc. es/DGDSI/Servicios/FirmaElectronica/Prestadores/relaPrestadores.htm.

12 *Real Decreto 1553/2005*, published December 24, 2005, regulates the issue of the national identity document and certificates of electronic signature. The first electronic DNI was issued March 16, 2006. By February 12, 2008 a total of 2,550,000 electronic DNIs had been issued. http://www.dnielectronico.es.

13 Criteria for accessibility applicable to the Web pages on Internet are defined internationally, by the Web Accessibility Initiative (WAI) of the World Wide Web Consortium. These guidelines contain the specifications of reference that allow Internet pages to be made accessible to persons with a handicap. The WAI sets three levels of accessibility: basic, medium and high, or A, AA and AAA. These standards are incorporated in Spain through UNE 139803:2004.

14 Interactive Digital Television (IDT) is projected as one of the most promising technological innovations of the future for e-administration access. In February 2008 the *Asociación Española de Usuarios de Telecomunicaciones y de la Sociedad de la Información* (Autelsi) gave one of its annual awards to the *Instituto Nacional de Tecnologías de la Comunicación* (INTECO) in the category "Best technological project in the Public Sector" for its development of the Platform of Public Services in TDT.

15 In the Report 2007 "Impact of corruption in different sectors and institutions" Spaniards gave the sector "political parties" a mark of 3.9 out of 5. http://www. transparencia.org.es/.

16 In the manifesto, these entities point out that it is unacceptable for Spain, along with Greece, to be one of the only two countries of the European Union with over one million inhabitants that does not have a specific law regulating the right to access to information.

17 Modes of participation are many, and the Web 2.0 presents the technical requirements for each. Participation based on voting has applications for the guarantee of confidentiality. The Spanish newspaper La Vanguardia printed an article in December 2007 about the Spanish software firm, Scytl. This company, with offices in Singapore and Washington, has come up with an application for voting at a distance. Its technology was selected by the County of Okaloosa, Florida, so that their citizens residing citizens could vote by Internet in the country´s presidential elections. La Vanguardia. "La e-democracia que viene". December 27, 2007, p. 60.

18 This law implements Directive 2003/98/CE on a Reuse of Public Sector Information.

19 Biblioteca Digital Hispánica (BDH), which gives free access to some of the major works of art of Spanish culture. Some 10,000 works can be consulted or downloaded, with manuscripts, books printed from the 15th century to the 19th century, etchings, drawings, posters, photographs and maps.

20 Established in accordance with Recommendation c (2003) 1422 of the European Commission, May 6, 2003.

21 These data may be inverted depending on the size of the municipality. For example, the larger municipalities and the entities such as *diputaciones, consejos* and *cabildos* may dedicate up to 89% of their expenses to computer resources, while the smaller municipalities may spend up to 63% on telecommunications.

22 As mentioned elsewhere in the article, the studies undertaken in 2005 and 2006 by the autonomous communities through the *Observatorio de la Sociedad de la Información* are incomplete, as in earlier editions some of the regional governments did not respond to the questionnaires, and furthermore, it is the regional governments themselves who gather and serve the data. For this reason, we opted to cite the results from entities we consider to be independent.

23 *eEspaña 2007. Informe Anual sobre el Desarrollo de la Sociedad de la Información*. Madrid: Fundación France Telecom España, 2007. The Orange Foundation and Capgemini came to an agreement to assess the e-Government in Spain, in view of the experience of Capgemini in its European studies: On line Availability of public services in Europe.

24 In recent years, different specific observatories have measured the adoption of information society in different regional spheres: *Observatorio Aragonés de la Sociedad de la Información*; *Observatorio de la Sociedad de la Información de Asturias*; *Observatorio de la Sociedad de la Información de Cantabria*; *Observatorio Regional de la Sociedad de la Información de Castilla y León*; *Observatorio para la Sociedad de la Información en Cataluña*; *Observatorio de las Tecnologías de la Información y la Comunicación de Galicia*; *Observatorio de la Sociedad de la Información en Navarra*; *Observatorio Valenciano de la Sociedad de la Información* and its offshoot, *Centro Valenciano para la Sociedad de la Información* (CEVALSI). [All consulted 21 August 2006]. There are also organisms pertaining to the central government (*Comisión de Estudio para el Desarrollo de la Sociedad de la Información*), and firms in the Telecommunications sector (annual reports by Telefónica, reports from the Orange Foundation and Fundación CTIC).

25 The EU-15 comprises: Austria, Belgium, Denmark, Finland, France, Germany, Greece, Ireland, Italy, Luxembourg, Netherlands, Portugal, Spain, Sweden, and United Kingdom. We distinguish this group of 15 countries by the greater per capita income with respect to other eastern European countries that joined the EU.

26 The INE survey of equipment and use of ICT in the home in 2007, revealed that 60.4% of homes have a computer; 39,2% have a wide band connection (10 points higher than in 2006), 81.2% have a regular telephone, 90.9% have a mobile telephone, and 22.8% of Spanish homes have Digital Interactive Television.

## References

Allen, A. B., Juillet, L., Paquet, G., & Roy, J. (2001). E-Governance and government online in Canada: Partnerships, people and prospects. *Government Information Quarterly*, 18(2), 93—104.

Alonso Peña, C., Fernández Gómez, R., & Yánez Morante, E. (2007). Consultation and update of property information by web services. *X Jornadas sobre Tecnologías de la Información para la Modernización de las Administraciones Públicas* [10th Congress on Information Technologies for Modernizing Administrations] (Gijón, 27-30 November).

Arsntein, S. (1971, April). Ladder of participation in the USA. *Journal of the Royal Town Planning Institute*, 176—182.

Bertot, J. C., & Jaeger, P. T. (2008). The e-Government paradox: Better customer service doesn't necessarily cost less. *Government Information Quarterly*, 25(2), 149—154.

Biko2. (2008). Informe de la usabilidad de los portales las Comunidades Autónomas [Report of usability of the Autonomous Community portals]. San Sebastián: Biko2.

BOE. (1992). Ley 30/1992, de 26 de noviembre, de Régimen Jurídico de las Administraciones y del Procedimiento Administrativo Común Retrieved Sept. 20, 2007 from http://www.boe.es/t/es/bases_datos/doc.php?coleccion=iberlex&id=1992/26318.

BOE. (1996). Real Decreto 263/1996, de 16 de febrero, por el que se regula la utilización de técnicas electrónicas, informáticas y telemáticas por la Administración General del Estado Retrieved April 4, 2008 from http://www.boe.es/g/es/bases_datos/doc.php?coleccion=iberlex&id=1996/04594.

BOE. (2001). Ley 24/2001, de 27 de diciembre, de Medidas Fiscales, Administrativas y del Orden Social Retrieved October 1, 2007 from http://www.boe.es/g/es/bases_datos/doc.php?coleccion=iberlex&id=2001/24965.

BOE. (2002). Ley 34/2002, de 11 de julio de Servicios de la Sociedad de la Información y del Comercio Electrónico Retrieved April 1, 2008 from http://www.boe.es/g/es/bases_datos/doc.php?coleccion=iberlex&id=2002/13758.

BOE. (2003a). Real Decreto 209/2003, de 21 de febrero, por el que se regulan los registros y las notificaciones telemáticas, así como la utilización de medios telemáticos para la sustitución de la aportación de certificados por los ciudadanos Retrieved April 1, 2008 from http://www.boe.es/g/es/bases_datos/doc.php?coleccion=iberlex&id=2003/0415.

BOE. (2003b). Ley 51/2003, de 2 de diciembre, de igualdad de oportunidades, no discriminación y accesibilidad universal de las personas con discapacidad. Retrieved April 1, 2008 from http://www.boe.es/g/es/bases_datos/doc.php?coleccion= iberlex&id=2003/22066.

BOE. (2003c). LEY 59/2003, de 19 de diciembre, de firma electrónica Retrieved April 1, 2008 from http://www.boe.es/g/es/bases_datos/doc.php?coleccion=iberlex&id= 2003/23399.

BOE. (2007a). Ley 11/2007, de 22 de junio, para el acceso electrónico de los ciudadanos a las Administraciones Públicas Retrieved October 1, 2007 from http://www.boe.es/g/es/bases_datos/doc.php?coleccion=iberlex&id=2007/12352.

BOE. (2007b). Real Decreto 1494/2007, de 12 de noviembre.Reglamento sobre las condiciones básicas para el acceso de las personas con discapacidad a las tecnologías, productos y servicios relacionados con la sociedad de la información y medios de comunicación social Retrieved April 1, 2008 from http://www.boe.es/ g/es/bases_datos/doc.php?coleccion=iberlex&id=2007/19968.

BOE. (2007c). LEY 37/2007, de 16 de noviembre, sobre reutilización de la información del sector público Retrieved April 1, 2008 from http://www.boe.es/g/es/bases_datos/doc.php?coleccion=iberlex&id=2007/19814.

BOE. (2007d). Ley 56/2007, de 28 de diciembre, de Medidas de Impulso de la Sociedad de la Información Retrieved April 1, 2008 from http://www.boe.es/g/es/bases_datos/doc.php?coleccion=iberlex&id=2007/2244.

Bustelo, C., & García-Morales, E. (2008). Electronic administration, records management, and Spain's electronic access to public services legislation. *El Profesional de la Información*, 17(1), 106—111.

Chain Navarro, C., & Muñoz Cañavate, A. (2004). Análisis comparativo regional del desarrollo de la Administración local española en Internet (1997-2002) Regional comparative analysis of the development of Spanish local administration on the Internet]. *Investigación Bibliotecológica*, 18(36), 96—116.

Chain Navarro, C., Muñoz Cañavate, A., & Más Bleda, A. (2008). La gestión de información en las sedes Web de los ayuntamientos españoles [[Information management on the websites of Spanish town halls]]. *Revista Española de Documentación Científica*, 31(4), 612—638.

Chain Navarro, C., Muñoz Cañavate, A., & Salido Martínez, V. (2008). LIS education and web services in the public sector: The case of Spain. *Libri*, 58(4), 246—256.

Chan, C. M. L., Lau, Y., & Pan, S. L. (2008). E-Government implementation: A macro analysis of Singapore's e-Government initiatives. *Government Information Quarterly*, 25(2), 239—255.

Chen, A., Pan, S. L., Zhang, J., Wei Huang, W., & Zhu, S. (2009). Managing e-government implementation in China: A process perspective. *Information & Management*, 46(4), 203—212.

CIS. (2006a). Centro de Investigaciones Sociológicas. Cuestionario 2604. Madrid: CIS. CIS. (2006b). Centro de Investigaciones Sociológicas. Cuestionario 2655. Madrid: CIS. Comisión Europea. (2006). Plan de Acción sobre administración electrónica i2010: Acelerar la administración electrónica en Europa en beneficio de todos COM 173 final. Retrieved February 21, 2008 from http://ec.europa.eu/information_society/activities/egovernment_research/doc/highlights/comm_pdf_com_2006_0173_f_es_acte.pdf.

Cornella, A. (1998). Information policies in Spain. *Government Information Quarterly*, 15 (2), 197—220.

eEspaña 2006. (2006). Informe anual sobre el desarrollo de la Sociedad de la Información en España 2006 [[Annual report on the development of Information Society in Spain 2006]]. Madrid: Fundación Orange France Telecom.

eEspaña 2008. (2008). Informe anual sobre el desarrollo de la Sociedad de la Información en España 2008 [[Annual report on the development of Information Society in Spain 2008]]. Madrid: Fundación Orange France Telecom.

Esteves, J. (2005). Análisis del desarrollo del gobierno electrónico municipal en España [[Analysis of the development of municipal electronic government]]. Madrid: Software AG.

Esteves, J., & Joseph, R. C. (2008). A comprehensive framework for the assessment of eGovernment projects. *Government Information Quarterly*, 25(1), 118—132.

Fundación Telefonica. (2004). La Sociedad de la Información en España 2004 [[Information Society in Spain 2004]]. Madrid: Telefonica.

Fundación Telefonica. (2006). La Sociedad de la Información en España 2006 [[Information Society in Spain 2006]]. Madrid: Telefonica.

Fundación Telefonica. (2007). La Sociedad de la Información en España 2007 [[Information Society in Spain 2007]]. Madrid: Telefonica.

Gauld, R., Gray, A., & McComb, S. (2009). How responsive is e-Government? Evidence from Australia and New Zealand. *Government Information Quarterly*, 26 (1), 69—74.

Gil-García, J. R., & Martínez-Moyano, I. J. (2007). Understanding the evolution of e-Government: The influence of systems of rules on public sector dynamics. *Government Information Quarterly*, 24(2), 266—290.

Groznik, A., & Trkman, P. (2009). Upstream supply chain management in e-Government: The case of Slovenia. *Government Information Quarterly*, 26(3), 459—467.

Gupta, M. P., & Jana, D. (2003). E-government evaluation: A framework and case study. *Government Information Quarterly*, 20(4), 365—387.

Hiller, J. S., & Bélanger, F. (2001). Privacy strategies for electronic government. In M. A. Abramson & G. E. Jeans (Eds.), *E-government 2001* (pp. 162—198). Lanham, MD: Rowman & Littlefield Publishers.

Holliday, I., & Yep, R. (2005). E-government in China. *Public Administration and Development*, 25(3), 239—249.

INFOAGE. (2005). Informe analítico de gestión y orientación de las tecnologías de la información y las comunicaciones (TIC's) en la AGE y estudio de la administración electrónica en Europa [[Analytical report on the management and orientation of information and communication technologies in Spain´s General Administration and study of e-administration in Europe]]. Madrid: ASTIC.

Jaeger, P. T., & Thompson, K. M. (2003). E-Government around the world: Lessons, challenges, and future directions. *Government Information Quarterly*, 20(4), 389—394. Koelher, W. (2002). Web page change and persistence—a four-year longitudinal study. J*ournal of the American Society for Information Science and Technology*, 53(2), 162—171.

Layne, K., & Lee, J. (2000). Developing fully functional e-government: A four stage model. *Government Information Quarterly*, 18(2), 122—136.

MAP (1995). Directorio de servicios de información electrónica de la Administración General del Estado [[Directory of e-Administration information services of the General Administration of Spain]]. Madrid: MAP.

MAP. (2006). Informe IRIA 2006. Las tecnologías de la información en las administraciones públicas [[IRIA Report 2006, Information technologies in the public administration]]. Madrid: MAP.

MAP. (2008). Informe IRIA 2008 Las Tecnologías de la Información en las Administraciones Públicas [[IRIA Report 2008, Information Technologies in the Public Administration]]. Madrid: MAP.

Martínez Domínguez, M., & García de la Paz, A. J. (2007, November 27-30). Plataforma individual del voto electrónico [[Individual platform for electronic voting]]. Ministerio de Administraciones Públicas (Public Administration Ministry).

Ministerio de la Presidencia. (2009). Informe REINA 2009. Las Tecnologías de la Información y de las Comunicaciones en la Administración del Estado [[REINA Report 2009, Information Technologies and Communications in the State Administration]]. Madrid: Ministerio de la Presidencia.

Mitra, R. K., & Gupta, M. P. (2008). A contextual perspective of performance assessment in e-Government: a study of Indian Police Administration. *Government Information Quarterly*, 25(2), 278—302.

Moulaison, H. L. (2004). The Minitel and France's legacy of democratic information access. *Government Information Quarterly*, 21(1), 99—107.

Muñoz Cañavate, A. (2003). La Administración General del Estado en Internet. Un estudio sobre la VI Legislatura [The General State Administration on Internet. A study of the 6th Legislature]. Badajoz: ICE.

Muñoz-Cañavate, A., & Chain-Navarro, C. (2004a). La Administración local española en Internet: estudio cuantitativo de la evolución de los sistemas de información Web de los ayuntamientos (1997-2002) [The Spanish local administration on the Internet: a qualitative study of the evolution of Web information systems for town halls]. *Ciencias de la Información*, 35(1), 43─55.

Muñoz-Cañavate, A., & Chain-Navarro, C. (2004b). The World Wide Web as an information system in Spain's Regional Administrations (1997-2000). *Government Information Quarterly*, 21(2), 198─218. OCDE. (2007). Economic Surveys: Spain 2007. : OECD Publishing.

Reventos, L. (2008, January 3). 2008, fin del BOE en papel [2008, end of BOE on paper] El País. Retrieved February 21, 2008 from http://www.elpais.com/articulo/portada/ 2008/fin/BOE/papel/elpeputeccib/20080103elpcibpor_2/Tes.

Software, A. G. (2005). Análisis del Desarrollo del Gobierno electrónico Municipal en España 2005 [Analysis of the Development of Municipal e-Government in Spain 2005]. Madrid: Software AG.